\documentclass{article}
\usepackage{titling}
\usepackage{bbm}
\usepackage{float}
\usepackage{color}
\usepackage{natbib}
\usepackage{graphicx}
\usepackage{amsfonts}
\usepackage{bookman}
\usepackage{tikz}
\usepackage{pgf}
\usepackage{url}
\usetikzlibrary{shapes}
\usetikzlibrary{calc}
\usetikzlibrary{positioning} 
\usetikzlibrary{decorations.pathreplacing}
\newcommand{\indep}{\perp \!\!\! \perp}
\usepackage{amsmath}

\usepackage[utf8]{inputenc}
\usepackage{fancyhdr}
\usepackage{hyperref}
\usepackage[letterpaper, margin=1.5in]{geometry}
\usepackage{colortbl,dcolumn}
\title{Learning and Testing Sub-groups with Heterogeneous Treatment Effects:\\ A Sequence of Two Studies}
\author{\href{https://rahulladhania.com}{Rahul Ladhania}\thanks{University of Pennsylvania}
 \href{http://www.heinz.cmu.edu/faculty-and-research/faculty-profiles/faculty-details/index.aspx?faculty_id=138}{ Amelia Haviland}\thanks{Carnegie Mellon University} \\ \href{http://healthpolicy.usc.edu/ListItem.aspx?ID=14}{ Neeraj Sood}\thanks{University of Southern California}  \href{http://www.stat.cmu.edu/people/faculty/edward-kennedy}{ Edward Kennedy$^2$} \href{https://www.hcp.med.harvard.edu/faculty/core/ateev-mehrotra-md}{Ateev Mehrotra}\thanks{Harvard Medical School}}
\date{}
\thanksmarkseries{arabic}
\begin{document}
\vspace{-12ex}
\maketitle
\vspace{-4ex}
\begin{abstract}
 There is strong interest in estimating how the magnitude of treatment effects of an intervention vary across sub-groups of the population of interest. In our paper, we propose a two-study approach to first propose and then test heterogeneous treatment effects. In Study 1, we use a large observational dataset to learn sub-groups with the most distinctive treatment-outcome relationships (`high/low-impact sub-groups'). We adopt a model-based recursive partitioning approach to propose the high/low impact sub-groups, and validate them by using sample-splitting. While the first study rules out noise, there is potential bias in our estimated heterogeneous treatment effects. Study 2 uses an experimental design, and here we classify our sample units based on sub-groups learned in Study 1. We then estimate treatment effects within each of the groups, thereby testing the causal hypotheses proposed in Study 1. Using patient claims data from the NBER MarketScan database, we apply our approach to estimate heterogeneous effects of a switch to a high-deductible health insurance plan on use of outpatient care by patients with a common chronic condition. We extend the method to non-parametrically learn the sub-groups in Study 1. We also compare the methods' performance to other state-of-the-art methods in the literature that make use only of the Study 2 data.
\end{abstract}
\linespread{2.0}
\section{Introduction}
There is strong interest in estimating how the magnitude of treatment effects of an intervention vary across different sub-groups of the population of interest. Specifically, in fields such as health-care (precision medicine, insurance plan design), there is value in understanding which patient/enrollee sub-groups will be high (or low) `responders' to a treatment regime(s), and the magnitude of response therein. In a number of settings, researchers are hobbled by a lack of theory to \textit{a priori} determine sub-groups of interest, a problem further aggravated by the relative high dimensionality of the co-variate space. In the statistical machine learning literature, recent studies have used a variety of approaches - tree and forest based methods (\textcolor{blue}{\cite{wager2018estimation}}; \textcolor{blue}{\cite{athey2016recursive}});  family of representation-learning based algorithms (\textcolor{blue}{\cite{shalit2016estimating}}) - to identify heterogeneous treatment effect sub-groups. Notably, these studies estimate heterogeneity in causal effects after a study with exogenous variation in treatment, i.e. experimental or observational settings with plausible unconfoundedness. When the sample size is  `large'  enough, these estimators have been demonstrated to  be consistent with well-understood asymptotic sampling distributions. These approaches, however, have limitations when sample size is a concern, which is often the case in experimental settings which, apart from being expensive (vis-à-vis observational settings) are often designed to have sufficient power to capture the average treatment effect. With a small sample size, lack of power is the key reason why it becomes difficult to rule out noise as the cause of observed heterogeneity in treatment effects. By chance alone, there is a high likelihood that some groups show larger effects, and cross-validation is a common strategy to rule out noise. Sample-splitting essentially worsens the problem by reducing the effective sample size available to learn these groups. Moreover, high-dimensional settings only serve to aggravate this problem.\\ \\
In our paper, we propose a two-study approach to mitigate this problem. We learn and test heterogeneous treatment effect sub-groups in a two-study setting. The first study uses data with endogenous treatment (observational setting \textbf{with} potential confounding) to learn and rule out noise as the cause of potential sub-groups. The second study uses data with exogenous variation in treatment to test for heterogeneous treatment effects in the proposed sub-groups. In most applied policy settings, observational data (e.g. patient claims data), while being of limited  use in causal inference due to valid concerns regarding endogeneity, are large and inexpensive, especially in comparison to experimental data. Our approach exploits the `largeness’ of the observational data to propose and validate heterogeneous association sub-groups, where the associations are demonstrated to be persistent, thus not induced by noise. Study 1, thus, is performed in an observational data setting, where the goal is to generate specific hypotheses about noise-free sub-groups that may be high/low impact. We define sub-groups based on a combination of the observed co-variates, and hypothesize them to be high (or low) impact based on `distinctiveness' of the estimated relationship in the sub-group between the outcome and treatment variable relative to the relationship in the pooled population. We do this using a  parametric and a non-parametric approach. In both the approaches, we adopt sample-splitting by randomly splitting our Study 1 data into training and validation data-sets, where we learn the groups on the training data-set, while ``validating" them on our validation data-set. In the parametric approach, we learn the sub-groups with distinctive relationships in Study 1 using a model-based recursive partitioning approach proposed in \textcolor{blue}{\cite{seibold2016model}} on our training data-set. To ensure the associations in the sub-groups we learn are not random noise, we  eliminate sub-groups which do not demonstrate `similar' associations in the training and validation datasets of the observational data. \\ \\
In the non-parametric approach, we run ensemble supervised machine learning methods separately on the treatment and control groups in our training data-set to estimate the predicted outcomes based on their co-variates (clinical and demographic features). We define the ``treatment-control difference" for each of the units in our training sample as the difference between the observed outcomes and the `counterfactual' ones we predict. We then fit a  decision tree with the ``treatment-control difference" as the outcome variable to learn the sub-groups. Analogous to our approach in the parametric method, we  eliminate sub-groups which do not demonstrate `similar' average ``treatment-control differences" in the training and validation datasets. \\ \\
Thus, in both the parametric and non-parametric approaches, the results of the first study is a set of sub-groups with observed heterogeneous `treatment-control' differences that are not due to noise. \\ \\ 
We then move on to the testing stage in Study 2, which is an experimental setting. We `transport' the partitioning rules learned in Study 1 as \textit{a priori} hypotheses, and classify Study 2 units based on their co-variates into these proposed sub-groups. We now use the exogenous variation in treatment that the experimental setting affords us to test for the hypothesized causal relationships in the sub-groups, allowing testing of our proposed hypotheses from Study 1.
\subsection{Generalizibility of inferences}
In our setup, we are essentially transporting causal hypotheses from one setting (observational) to another (experimental). There is a large body of work which has been done in the space of generalizing/transporting inferences across different settings [\textcolor{blue}{\cite{cole2010generalizing}; \cite{tipton2013improving}}]. The  key challenge in generalizing/transporting causal inferences from one setting to specified targets is the difference in distributions of baseline covariates between participants in the settings. While a sound experimental design is the best bet in dealing with these challenges, studies have proposed statistical approaches to deal with them. In  \textcolor{blue}{\cite{pearl2014external}}, the authors explicitly lay out graph-based criterion to determine when transportability is feasible, and when it is, to identify what experimental and observational
findings need be obtained and combined from the two populations to ensure bias-free transport. In our setting, we `transport' hypotheses from an observational setting to an experimental setting, with the explicit assumption about overlap in distributions of covariates.  \\ \\
To our knowledge, this is the first study that proposes a novel two-study approach towards estimating heterogeneity in treatment effects. In exploiting the `largeness' (despite potential unobserved confounding) of the observational data, our approach better informs heterogeneity estimation in the experimental dataset by proposing noise-free/stable a priori hypotheses, a feature we did not see in other studies. This is in contrast to the dominant body of work in the machine learning literature, which does the estimation in an experimental or observational (with a no unobserved confounding assumption) settings, and are often under-powered in empirical settings to rule out noise. We also compare our approach to causal trees [\textcolor{blue}{\cite{athey2016recursive}}] and find that our method out-performs that method at identifying stable heterogeneous treatment effect sub-groups.\\ \\
 The rest of the paper is structured as follows. We describe the two-study parametric and non-parametric approaches in Section \ref{gen_inst}, and describe how we apply the approach to patient claims data in Section \ref{application}. We  lay out the empirical problem setup in Section \ref{problem_setup}, and define how we compare our method's performance to existing methods for estimating heterogeneity in Section \ref{experiments}. In Section \ref{results}, we summarize the empirical results, and we discuss the implications and ongoing work in Section \ref{discussion}. We conclude in Section \ref{conclusion}.

\section{Method}
\label{gen_inst}
\subsection{Study 1A: Proposing sub-groups using observational data - parametric approach}
\label{studyone}
As described earlier, one key aspect of observational data in a number of applied policy settings is their relatively large sample size, and in Study 1, our aim is to learn plausibly heterogeneous treatment effect sub-groups where the different sized associations (magnitudes of the treatment-control difference) are not due to noise. We use a sample-splitting approach to identify the noise-free sub-groups - we split the data into `training' and 'validation' sets, and adopt a model-based recursive partitioning approach proposed by \textcolor{blue}{\cite{zeileis2008model}} and  \textcolor{blue}{\cite{seibold2016model}} to first learn the sub-groups on the `training' set. We then use the  `validation' set to identify those sub-groups where the effect persists which we will call ``stable" heterogeneous sub-groups. We emphasize that Study 1 is a hypothesis-generation stage - owing to the potential presence of confounding, we do not claim the identified sub-groups are causal, and all references to treatment effects in this section imply only potential causality which needs to be tested further.

\subsubsection{Generate sub-groups using model-based recursive partitioning}
\label{studyonepointone}
As a first step, we specify a parametric model to estimate the relationship between the treatment and outcome variables. We start with a base linear regression model describing the conditional distribution of the outcome, denoted by $Y$, as a function of ``treatment'', denoted by a binary or continuous treatment variable (with linear dose response assumption) $A$, and other co-variates, $\tilde{X}$: 
\begin{equation}
    Y = \beta_0 + \beta_1  A + \tilde{\beta_3} \tilde{X}+ \epsilon 
    \label{outcomemodel}
\end{equation} 

While there may be heterogeneous sub-groups that differ in their treatment effects, $\beta_1$ here only reflects the average treatment effect. We describe sub-groups as a partition $\{\mathcal{B}_b\} (b=1,\dots, B)$ of all units in the sample $i=1,\dots, N$. Here the partition $\{\mathcal{B}_b\}$ is defined by $J$ partitioning variables $\mathbb{Z} = (Z_1,Z_2,\dots,Z_J) \in \mathcal{Z}$. 

These partitioning variables are characteristics of our units that we posit would determine sub-groups with heterogeneous response to treatment. Without any \textit{a priori} knowledge about the partitions, we want to estimate the functions $\beta_0(z)$ and $\beta_1(z)$ using model-based recursive partitioning as applied in \textcolor{blue}{\cite{seibold2016model}}. The main idea here is to detect parameter instabilities - non-constant intercepts ($\beta_0$) and `treatment effects' (captured by $\beta_1$) - in the specified model by evaluating  the partial score functions with respect to $\beta_0$ and $\beta_1$: $$\psi_{\beta_0}(
(Y,A,X),\vartheta)= \frac{\delta\Psi((Y,A,X),\vartheta)}{\delta\beta_0}$$ and $$\psi_{\beta_1}(
(Y,A,X),\vartheta) = \frac{\delta\Psi((Y,A,X),\vartheta)}{\delta\beta_1}$$ Here $\vartheta$ denotes the model parameters. If the model parameters are constant and do not depend on the partitioning variables, then the partial score functions $\psi_{\beta_0}((Y,A,X),\vartheta)$ and $\psi_{\beta_1}((Y,A,X),\vartheta)$ are independent of $\mathbb{Z}$. Consequently, parameter instability corresponds to a correlation between either of the partial score functions and at least one of the partitioning variables. In order to formally detect deviations from independence between the partial score functions and the partitioning variables, model-based recursive partitioning utilizes independence tests. If we can reject at least one of the $2 \times J$ null hypotheses for the pooled model at a pre-specified nominal level, model-based recursive partitioning selects the partitioning variable $Z_j$
associated with the highest correlation to any of the partial score functions. On finding an optimal cut-point $Z_j* < \mu$ using a suitable criterion, we split the sample into two subgroups according $Z_j* < \mu$.  For both subgroups, we estimate two separate models with parameters, $\hat{\vartheta}(1)$  and $\hat{\vartheta}(2)$ respectively, obtain the corresponding partial score functions, and test the independence hypotheses. If we find deviations from independence, we in turn estimate a cut-point in the most highly associated partitioning variable, and split again. The procedure of testing independence of partial score functions and partitioning variables is repeated recursively until deviations from independence can no longer be detected (or based on other tuning parameters, such as minimum cluster size, to ensure that our sub-groups are large enough for testing stability of the associations in second study). 

\subsubsection{Validation set to rule out `noisy sub-groups'} 
\label{studyonepointtwo}
Arguably, several different criteria could be used to classify learned sub-groups as stable or noisy, and the choice may be application-dependent. One potential way could be to compare the coefficient estimates in the sub-groups in the training and validation sets, and define as stable those clusters which have similar estimates across the two sets. Alternatively, some contexts might be less concerned with matching the magnitude of the estimates in the two sets, and more with preserving the rank-order of the estimates across the sets. In our study, we apply the partitioning rules learned from running the recursive partitioning algorithm on the `training set' to data in the `validation' set, and estimate the $\beta_{1,val}(n)$ value for each resulting sub-group (denoted by $n$) in the validation set. We then compare these estimates to the $\beta_{1,train}(n)$ estimates for that same sub-group we learnt using the training data. We define `stable high-impact' sub-groups as sub-groups (denoted by $n$) for which $\beta_{1,train}(n)$ and  $\beta_{1,val}(n)$  are both significantly different from zero at the 0.05 level \textbf{and} are in the same direction \textbf{and} fail to reject the null when we test if the difference $\beta_{1,train}(n) - \beta_{1,val}(n)$ is significantly different from zero at the 0.05 level. 

\subsection{Study 1B: Proposing sub-groups using observational data - Non-parametric approach} 
\label{studyonea}
One of the limitations of the discussed approach in section above is that it relies extensively on a particular parametric model specification (Equation \ref{outcomemodel}). In a number of settings, there is reason to doubt whether the parametric model is valid. Further, in settings with a large co-variate space or in continuous treatment cases (linear dose response assumption needed), a simple parametric model as we use can be problematic. As the sub-group identification builds off on the specified model, the parametric approach also assumes similar treatment-outcome parametric relationships across sub-groups, which need not necessarily be true. Moreover, a non-parametric model to estimate the treatment-control differences could more exhaustively capture the the effect of observed confounders within the sub-groups (as in this case, we are including all the co-variates to predict the `counterfactual' outcomes), compared to the linear model used in Equation \ref{outcomemodel}. To address these concerns, we propose a non-parametric approach for learning and validating the sub-groups in Study 1. \\ \\
We relax the parametric model assumption in Study 1 to estimate the sub-groups non-parametrically using the following approach. Our objective from this step is to learn ``stable" sub-groups . In addition to addressing concerns over model specification, this approach does more to address potential observed confounding in Study 1 sub-group identification. This method is a plug-in version of the `counterfactual' and sub-group estimation strategy adopted in \textcolor{blue}{\cite{van2014targeted}} and \textcolor{blue}{\cite{foster2011subgroup}}. \\ \\
We devise the non-parametric learning \& estimation as a three-step procedure. We state these three steps below, and cover them in detail in Section \ref{threesteps}.
\begin{enumerate}
    \item  We non-parametrically estimate the ``counterfactual" outcome for each individual in our sample and then use that to estimate the \textbf{{treatment-control difference (TCD)}} for each individual. 
    \item We then partition individuals into sub-groups based on their TCD.
    \item Finally, we validate the `stability' of the learnt sub-groups. 
\end{enumerate}

\subsection{Defining the Treatment-Control Difference parameter}
\label{threesteps}
\begin{figure}[tbhp]
\label{figure_nonparametriclearning_1}
\resizebox{\columnwidth}{!}{ 
\begin{tikzpicture}
\draw (14,-1.5) node [draw, align= center]{\LARGE{\textbf{OBSERVATIONAL DATA}}};
\draw (2,0) -- (26,0) -- (26,14) -- (2,14) -- (2,0);
\draw[] (10,-0.5) -- (10,15);
\draw[] (18,-0.5) -- (18,15);
\draw (6,7) node [rectangle, draw, align = center ]{\LARGE{Learn} \\  \LARGE{$\hat{\gamma}_{\text{trn}}(X)$}};
\node (a) at (6, 15) {};
\node (b) at (13.5, 15) {};
\draw[->] (a)  to [out=90,in=90] (b);
\draw (14,10.5) node [rectangle, draw, align = center ]{\LARGE{Predict} \\  \LARGE{$\hat{\gamma}_{\text{pred}}(X)$}};
\draw (14,7) node [rectangle, draw, align = center ]{\LARGE{Fit tree based on $\hat{\gamma}_{\text{pred}}(X)$}  \\ \LARGE{to learn sub-groups}};
\draw (14,3.5) node [rectangle, draw, align = center ]{\LARGE{Calculate mean  $\hat{\gamma}_{\text{pred}}(n)$} \\ \LARGE{in each sub-group $n$}};
\node (c) at (14.5, 15) {};
\node (d) at (22, 15) {};
\draw[->] (c)  to [out=90,in=90] (d);
\draw (22,10.5) node [rectangle, draw, align = center ]{\LARGE{Apply learnt tree}  \\ \LARGE{rules to assign units}\\ \LARGE{to sub-groups}};
\draw (22,7) node [rectangle, draw, align = center ]{\LARGE{Learn} \\ \LARGE{$\hat{\gamma}_{\text{val}}(X)$}};
\draw (22,3.5) node [rectangle, draw, align = center ]{\LARGE{Calculate mean  $\hat{\gamma}_{\text{val}}(n)$}  \\ \LARGE{in each sub-group $n$}};
\end{tikzpicture}
}
\caption{Non-parametric learning and validation of sub-groups}
\end{figure}
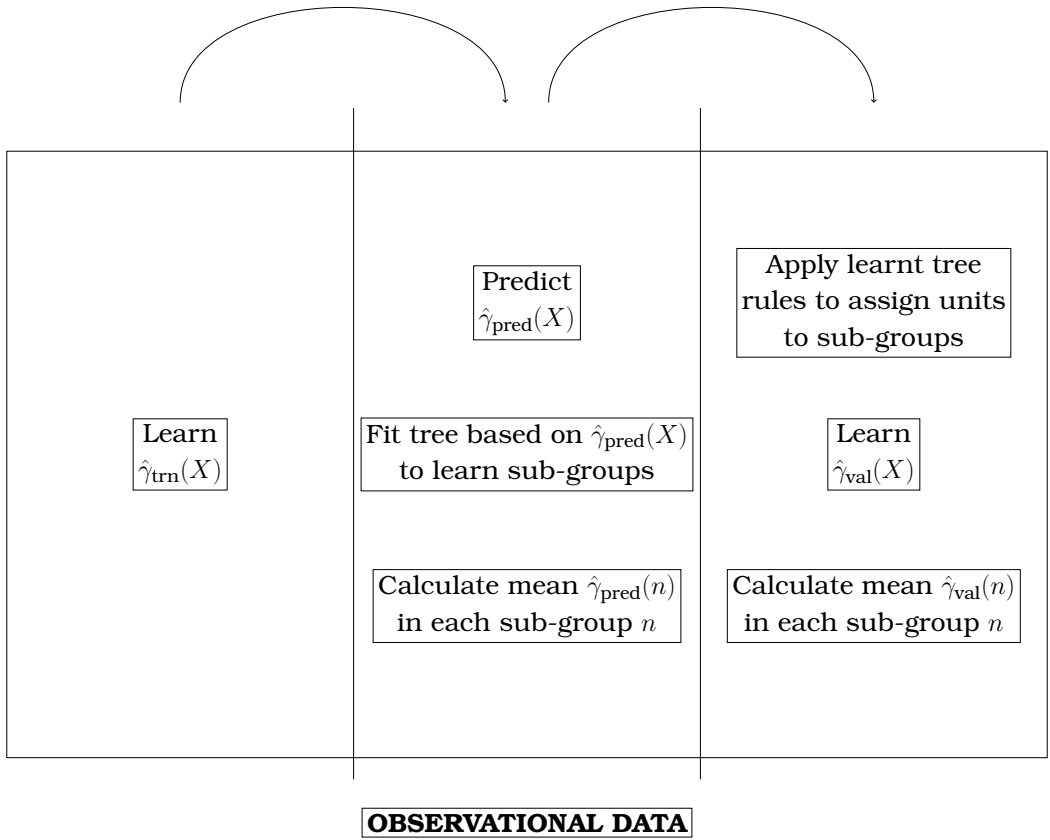

In a dichotomous treatment setting ($A \in \{0,1\}$), for each unit in the `training' data, we define the conditional treatment effect ${\gamma}$ as:
$${\gamma}(X)= E(Y^1-Y^0|X)$$
$$\implies {\gamma}(X)= E(Y^1|X)-E(Y^0|X)$$ 
Assuming conditional exogeneity/randomization $[E(Y^a|X) =E(Y^a|X,A=a)]$, we have:
$$\implies {\gamma}(X)= E(Y^1|X,A=1)-E(Y^0|X,A=0)$$ 
By consistency $[Y=AY^1+(1-A)Y^0]$, we have:
$$\implies {\gamma}(X)= E(Y|X,A=1)-E(Y|X,A=0)$$ 
However, in an observational setting, the first assumption (unconfoundedness) is a problematic one. $\gamma(X)$ in a setting where this assumption does not hold is \textbf{not} the conditional treatment effect parameter, and any estimation of the parameter should not be interpreted thus. For the purpose of hypothesis generation, however, which is the objective of our Study 1, we call ${\gamma}(X)$  the {\textbf{treatment-control difference (TCD)}}.
\\ \\
We follow these steps (depicted in Figure \ref{figure_nonparametriclearning_1}):
\begin{enumerate}
\item We randomly split our Study 1 observational data into three equally sized `training', `prediction' and `validation' sets.

\item  \textbf{Treatment-Control Difference (TCD)}: On `training' data, we estimate $E(Y|X,A=1)$ and $E(Y|X,A=0)$ separately on the treatment and control groups respectively to get the following estimators:
$$\hat{\mu}_1(X)= \hat{E}(Y|X,A=1)$$ 
$$\hat{\mu}_0(X)= \hat{E}(Y|X,A=0)$$ 
For every treatment unit $i$, we have: 
$$\hat{\gamma}(X_i)= y_1(X_i)- \hat{\mu}_0(X_i)$$ 
where $y_1(X_i)$ is the observed outcome of the treatment unit $i$. \\
For every control unit $j$, we have:
$$\hat{\gamma}(X_j)= \hat{\mu}_1(X_j) - y_0(X_j)$$ 
where $y_0(X_j)$ is the observed outcome of the control unit.

\item  \textbf{Estimation of ``counterfactual" outcomes on training data}: We estimate ${\mu}_0(X)$ and ${\mu}_1(X)$ on the training set to give us the following estimates: $\hat{\mu}_{\text{0,trn}}(X)$ and $\hat{\mu}_{\text{1,trn}}(X)$. 

We estimate each of the estimates by running two random forest models (one on control units, and one on treatment units) on the \textbf{training} subset of the observational data. We tune the hyper-parameters of these random forests (node size, tree depth, number of variables) using a cross-validation approach.

\item \textbf{Estimation of ``counter-factual" outcomes on validation data}: Similarly, we run two random forest models to estimate ${\mu}_0(X)$ and ${\mu}_1(X)$ on the validation set to give us  $\hat{\mu}_{0,val}(X)$, and $\hat{\mu}_{1,val}(X)$. 

\item \label{step:nonparamestimation} {Estimation of TCD}: Using the above, we calculate the treatment-control difference for every unit $i$ in the training set ($N_{\text{trn}}$) by using this formula:

For every treatment unit, we have:
$$\hat{\gamma}_{\text{trn}}(X_i)= y(X_i)-  \hat{\mu}_{\text{0,trn}}(X_i)$$ 
where $y(X_i)$ is the observed outcome of the treatment unit. \\
For every control unit $j$, we have:
$$\hat{\gamma}_{\text{trn}}(X_j)= \hat{\mu}_{\text{1,trn}}(X_j) - y(X_j)$$ 
where $y(X_j)$ is the observed outcome of the control unit.

We do the same for every unit in the validation set ($N_{val}$).
Treatment unit $i$:
$$\hat{\gamma}_{\text{val}}(X_i)= y(X_i)- \hat{\mu}_{\text{0,val}}(X_i)$$ 
Control unit $j$:
$$\hat{\gamma}_{\text{val}}(X_j)= \hat{\mu}_{\text{1,val}}(X_j) - y(X_j)$$

\item \label{learntree} \textbf{Learning sub-groups}: We now want to learn the partitioning rules to be able to apply them to the Study 2 experimental setting. We learn them by fitting a decision tree (CART) to the `prediction' data. Here we use with $\hat{\gamma}_{\text{pred}}(X)$ as the outcome variable for the CART and the co-variates $X$ as the partitioning variables, where $\hat{\gamma}_{\text{pred}}(X)$ is predicted using the forest models we fit on the `training' set. We use the variable vector $X$ to construct the tree. We use a cross-validation approach to avoid over-fitting. Thus, each of the terminal nodes $n$ in the learnt partitioning tree is a proposed sub-group, with $\hat{\gamma}_{\text{pred}}(n)$ as the average treatment-control difference for each sub-group $n$.

$$ \hat{\gamma}_{\text{pred}}(n) = \frac{\sum_{i=1}^{N_{\text{pred}}} \mathbbm{1}_{i \in n} \hat{\gamma}_{\text{pred}}(X_i)}{\sum_{i=1}^{N_{\text{pred}}} \mathbbm{1}_{i \in n}}$$

\item  We now apply the partitioning rules learned in Step \ref{learntree} to the validation set (note: we have already estimated the TCD $\hat{\gamma}_{\text{val}}(X)$ for each unit of the 
'validation set' in Step \ref{step:nonparamestimation}). Thus, for each of the terminal nodes $n$ in the validation set, we can then calculate the average $\hat{\gamma}_{\text{val}}(n)$

$$ \hat{\gamma}_{val}(n) = \frac{\sum_{j=1}^{N_{val}} \mathbbm{1}_{j \in n} \hat{\gamma}_{val}(X_j)}{\sum_{j=1}^{N_{val}} \mathbbm{1}_{j \in n}}$$

\item \textbf{Identifying `stable' sub-groups}: Analogous to the parametric setting in Section \ref{studyonepointone}, we define a sub-group $n$ as a ``stable" proposed high-impact group if $\hat{\gamma}_{\text{pred}}(n)$ and $\hat{\gamma}_{\text{val}}(n)$ are both significantly different from zero at the 0.05 level and are in the same direction; \textbf{and} if we fail to reject the null when we test if  the difference  $\hat{\gamma}_{pred}(n) - \hat{\gamma}_{val}(n)$ is significantly different from zero at the 0.05 level. 
We define a sub-group $n$ as a ``stable" proposed low-impact group if we fail to reject the null that $\hat{\gamma}_{pred}(n)$ and $\hat{\gamma}_{val}(n)$ are both significantly different from zero at the 0.05 level. We use boostrapping to estimate the standard errors of the mean TCD within each learnt subgroup
\end{enumerate}

\subsection{Study 2: Testing sub-groups in experimental setting}
\label{studytwo}
As noted earlier, in Study 1 we cannot conclude causality because of the presence of potential unobserved confounding in the observational data. This brings us to our Study 2 which is run in an experimental setting. We classify data (of a similar co-variate structure as the observational data from Study 1) from a setting with exogenous variation (experimental or quasi-experimental setup) into the sub-groups we learned and validated in Study 1. We then estimate the causal responses to treatment for each of these sub-groups. Depending on the nature of the outcome variable and the experimental setting, we can apply appropriate outcome modeling approaches based on Study 2 design - simple regression models for a randomized control trial or traditional difference-in-differences models or doubly robust semi-parametric estimation models - to estimate the effects. 

\section{Application to Patient Claims Data}
\label{application}
As a proof of concept, we apply the two-study approach to learn and test sub-groups with heterogeneous effects of a switch to high-deductible health plans (HDHPs) on use of outpatient care by 18-64 year old high-cholesterol patients. There has been extensive research on how high-deductible health plans (HDHPs) impact use of healthcare. In \textcolor{blue}{\cite{haviland2016consumer}}, the authors find that healthcare spending is reduced for enrollees in firms offering consumer directed health plans (CDHPs), and the reduction was driven by spending decreases in outpatient care and pharmaceuticals. In \textcolor{blue}{\cite{brot2017does}}, the authors find that enrollees at a firm that switched from an insurance plan that provided free health care to a high-deductible plan, reduced quantities across the spectrum of health care services, including potentially valuable care (e.g., preventive services) and potentially wasteful care (e.g., imaging services). They also find that these spending reductions came in large part from well-off and predictably sick consumers facing reasonably low yearly out-of-pocket maximums. In \textcolor{blue}{\cite{huckfeldt2015patient}}, the authors investigate how CDHP enrollees change use of pharmaceuticals for chronic diseases, and find reduction in utilization, regardless of whether the pharmaceuticals were exempt from the deductible. While several studies have focused on the average effect of a switch to HDHPs on use of care, understanding how the response differs across different patient sub-groups is an important policy question. There might be sub-groups of patients who might be harmed by enrollment in CDHPs due to forgoing needed care, or sub-groups for whom a CDHP switch might not a significant impact on health use. To the best of our knowledge, any assessment of heterogeneous responses in these studies was done ex-post, as opposed to testing for any specific a priori hypotheses.The problem gets understandably compounded by the largeness of the co-variate space (clinical and demographic characteristics) and and lack of theory for fine-grained sub-groups. Statistical machine learning techniques can be appropriate in situations such as this. \\ \\
In our study, we use  patient claims data from The Truven Health MarketScan\textsuperscript{\textregistered} Research Databases from two years (2010-2011). The MarketScan databases reflect the health-care experience of employees and dependents covered by the health benefit programs of large employers. These claims data are collected from approximately 100 different insurance companies. These data represent the medical experience of insured employees and their dependents for active employees, early retirees, COBRA continuees and Medicare-eligible retirees with employer-provided Medicare Supplemental plans. They capture person-specific clinical utilization, expenditures, and enrollment across inpatient, outpatient, prescription drug, and carve-out services. \\ \\
We exploit two features of this dataset which allow us to apply our two-study approach. First, one of the employers in our two-year time period switches completely to  high-deductible health insurance plans in the second year (2012) from a suite of mostly low-deductible health plan offerings in the first year (2011). This lays the premise for Study 2 by providing us a quasi-experimental setup where the switch to a high-deductible plan is an exogenous cost shock from an enrollee's perspective. Our sample has ($\textbf{N} \sim \textbf{8500}$) enrollees who meet our sample inclusion criteria in the full-replacement firm.  Second, our first year of data (2011) provides us the framework for Study 1, as we observe enrollees ($\textbf{N} \sim \textbf{130,000}$) meeting the sample inclusion criteria in a diversity of health insurance plans offered by multiple employers, with considerable variation in costs. We use this variation to learn the heterogeneous effect sub-groups. 

\section{Problem Setup}
\label{problem_setup}
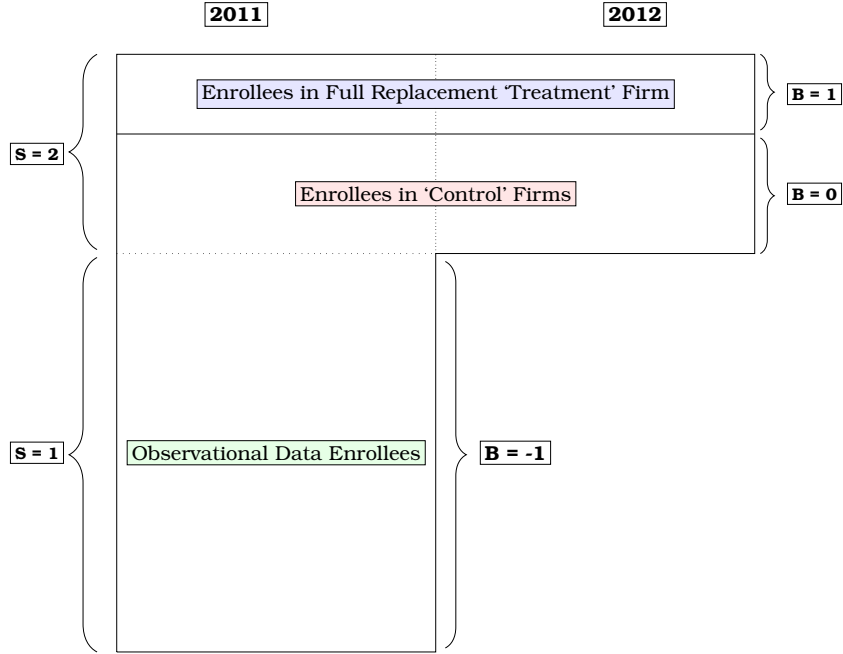
\begin{figure}[tbhp]
\centering
\resizebox{0.8\columnwidth}{!}{ 
\begin{tikzpicture}
\draw (2,0) -- (10,0) -- (10,10) -- (18,10) -- (18,15) -- (2,15) -- (2,0);
\draw[loosely dotted] (2,10) -- (10,10);
\draw[dotted] (10,10) -- (10,15);
\draw (5,16) node[rectangle, draw]{\Large{\textbf{2011}}};
\draw (15,16) node[rectangle,draw]{\Large{\textbf{2012}}};
\draw [decorate,decoration={brace,amplitude=20pt},xshift=0.2pt,yshift=0pt] (1.5,0) -- (1.5,9.9);
\draw (0,5) node [rectangle, draw]{\large{\textbf{S = 1}}};
\draw [decorate,decoration={brace,amplitude=20pt},xshift=0.2pt,yshift=0pt] (1.5,10.1) -- (1.5,15);
\draw (0,12.5) node [rectangle, draw]{\large{\textbf{S = 2}}};
\draw[] (2,13) -- (18,13);
\draw (6,5) node[fill=green!10,draw]{\Large{{Observational Data Enrollees}}};
\draw (12,5) node [rectangle, draw]{\Large{\textbf{B = -1}}};
\draw (10,11.5) node[fill=red!10,draw]{\Large{{Enrollees in `Control' Firms}}};
\draw (19.5,11.5) node [rectangle, draw]{\large{\textbf{B = 0}}};
\draw (10,14) node[fill=blue!10,draw]{\Large{{Enrollees in Full Replacement `Treatment' Firm }}};
\draw (19.5,14) node [rectangle, draw]{\large{\textbf{B = 1}}};
\draw [decorate,decoration={brace,amplitude=15pt,mirror,raise=4pt},yshift=0pt]
(10.1,0.1) -- (10.1,9.8) ;
\draw [decorate,decoration={brace,amplitude=10pt,mirror,raise=4pt},yshift=0pt]
(18,10) -- (18,12.9) ;
\draw [decorate,decoration={brace,amplitude=10pt,mirror,raise=4pt},yshift=0pt]
(18,13.1) -- (18,15) ;
\end{tikzpicture}
}
\caption{Two-Study Data Setup}
\label{data_viz}
\end{figure}

Figure \ref{data_viz} depicts the data setup for the two-study approach. Let $S \in \{1,2\}$ indicate the study state (1 or 2), where $S=1$ is Study 1 (observational setup), and $S=2$ be the Study 2 (experimental setup). Let $A_s$ indicate value of treatment variable (continuous) in any study $s$. \\ \\
In this dataset, as we have a fixed cost shock (for the full-replacement employer) in Study 2, we re-define  the treatment variable to assume a discrete value. We define $B$ as the indicator for treatment, where $t$ is the threshold value of the cost variable. Then,  
$$B=\begin{cases}
    1 & \text{if } A_2 \ge t \ge A_1 \\
    0 & \text{if } A_1, A_2 < t \\
    -1 & \text{if  otherwise }
  \end{cases}$$
Here, $B=1$ indicates enrollees in the full-replacement firm which are subject to the exogenous cost-shock, while $B=0$ indicates enrollees in firms which do not see any change in the cost shock over the two years (enrollees in these firms serve as `controls' for our `treatment' firm enrollees). $B=-1$ for enrollees in $S=1$. \\ 
\newpage We use Study 1 ($S=1$) to learn $\nu=f(X)$,  which indicates learned sub-groups from Study 1 and is defined as a function of the co-variates (partitioning variables $X$). \\ \\
For Study 2 ($S=2$), we define the Conditional Average Treatment Effect  (CATE) parameter to be estimated in each of the sub-groups $\nu$ as:
\begin{equation}
\text{CATE} = E[Y_2^{b=1}-Y_2^{b=0}\mid \nu, S=2] = \gamma(\nu)
\label{equation_cate}
\end{equation} 
Under   the following assumptions:
\begin{itemize} 
\item Ignorability
$$ B \indep Y_2^b \mid \nu, S= 2 $$ 
\item Consistency
$$ Y_2 = BY_2^1 + (1-B)Y_2^0$$ 
\end{itemize}
We use Equation \ref{equation_cate} :
$$ \gamma(\nu) =E[Y_2^{b=1}\mid \nu, S=2] - E[Y_2^{b=0}\mid \nu, S=2] $$ 
Using ignorability, this is equal to:
$$
\gamma(\nu)= E[Y_2^{b=1}\mid \nu, S=2, B=1] - E[Y_2^{b=0}\mid \nu, S=2, B=0]
$$
Then, using consistency, we have:
\begin{equation}
\gamma(\nu)= E[Y_2\mid \nu, S=2, B=1] - E[Y_2\mid \nu, S=2, B=0]
\end{equation}

\subsection{Sample Selection}
\label{sampleselection}
For our study, we focus our analysis on 18-64 year old adults who have a chronic condition - high cholesterol (hyperlipidemia). For patients with a chronic condition, there is a concern that reduction in health care use could be sub-optimal and welfare reducing. A sustained engagement with the health-care provider - captured by outpatient visits - is presumed to be optimal for this patient group. Further, having a more homogeneous study population - those with this particular chronic condition - may reduce issues of selection into plans. For our sample, we select 18-64 year old enrollees (or spouse) who \begin{itemize}
\item are continuously enrolled through the same employer for 24 months (2011-12), \textbf{and}
\item have at least one claim in the outpatient services file for the first six months of 2011, with any of the four diagnostic codes classified for hyperlipidemia.
\end{itemize}
We note that from an empirical perspective, as we have access to claims data alone, for constructing enrollees' clinical profiles, we are limited to enrollees that have at least one claim in the outpatient services.

\subsection{Study 1A: Learning sub-groups - parametric approach}
\label{parametricimplementation}
\subsubsection{Defining `Outcome' and `Treatment' measures}
\label{outcomeandtreatment}
We estimate the heterogeneous effects of a switch to a high-deductible health insurance plan on patients' use of care. In our setting, we define care to specifically mean outpatient (OP) care, as measured by \textbf{number of OP visits}.  We define the treatment variable as the cost exposure enrollees face for an outpatient visit. We posit that the effect of the switch to a high-deductible plan is through exposure to the out-of-pocket cost shock that enrollees face due to the switch. To define the out-of-pocket cost exposure, we construct a health-plan level measure which is similar to an actuarial value for OP visits in particular. This is needed as using the raw out-of-pocket costs that patients face (sum of copay/coinsurance and deductible) across visits does not capture the non-linearity in prices patients face over time (due to plans' cost-sharing features) and are also endogenous to the enrollee. Moreover for comparing across plans, any such measure also needs to account for heterogeneity in plans' enrollees - to mitigate the issue of enrollees selecting into plans  based on information private to the enrollees. We define the measure in the appendix. We note that our treatment variable (cost exposure) is a continuous variable, and we assume a linear dose response relationship in the parametric model. \\ \\

Our pooled model is the following simple linear regression model:
\begin{equation}
    \hspace{-3ex}\log(\text{OP Visits})_{i} = \beta_0 + \beta_1 \times \text{Cost Exposure}_{i}  + \epsilon_i
    \label{outcomemodel_data}
\end{equation} 

\subsubsection{Partitioning Variables}
\label{partitioning}
In the case of our study, we have patient demographic features - age and gender - and a summary of clinical features we obtain from the inpatient admissions, outpatient services, and outpatient drug files. We use patients' 2011 claims data to construct their clinical and demographic profiles. Below are the clinical features we consider in the model-based recursive partitioning algorithm:
\begin{itemize}
    \item Number of unique therapeutic drug classes (range of values)
    \item Indicator of whether patient was ever admitted to the hospital in the time-period of the observational data
    \item Indicators of whether patient had an OP service claim ever for each of the top ten most-frequently-occurring `Major Diagnostic Categories' (MDCs) 
	\item Indicators of whether patient had an OP drug claim ever for each of the top twenty most-frequently-occurring therapeutic drug classes
	\end{itemize}

We also define a categorical variable that categorizes patients into three bins based on their overall health-care expenses (total payments inclusive of out-of-pocket payments and payments made by insurer). Patients with total payments less than 2500 USD were in the `low strata', with payments between 2500-6000 USD were in `medium strata', and those with payments higher than 6000 USD were in the `high strata'. We note that feature selection is a complex task open to subjectivity. We employ the following procedure to inform the feature selection procedure. We ran three combinations of the aforementioned partitioning variables (in increasing levels of granularity) with varying levels of the minimum cluster size (1000,1500,2000, 3000), and used a cross-validation approach to select the feature set and cluster size combination using `validation' set MSE/AIC as the criterion.

\subsubsection{Model-based Recursive Partitioning}
As described in Section \ref{studyone}, we apply the model-based recursive partitioning algorithm to the 2011 claims data for patients in our sample. We exclude patients from the full-replacement firm while forming the groups, and also exclude patients from three `comparison' firms - these are firms for which we do not see any change in the nature of plans (similarly priced) offered over 2010-11. 

\subsection{Study 1B: Learning sub-groups - non-parametric approach}
\label{nonparametricimplementation}
\begin{figure}
\label{costshock}
\centering
\includegraphics[width=0.9\linewidth]{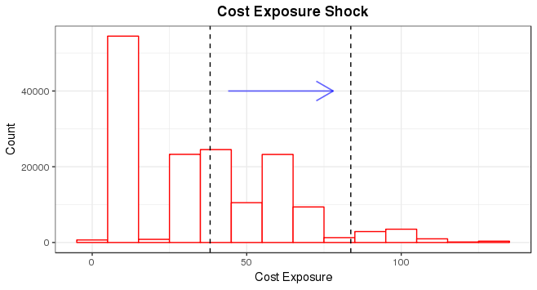}
\caption{Cost Exposure Shock in Study 2}
\end{figure}

We now use the approach as described in Section \ref{studyonea}. We use the same 2011 claims data and the inclusion criteria and the partitioning variables as described in Sections \ref{sampleselection} and \ref{parametricimplementation}. A major difference between the two approaches is how we define our treatment variable. In the parametric approach, the weighted cost-exposure was considered as a continuous treatment variable, with a linear dose-response assumption. For our non-parametric method, we need to explicitly define our treatment and control individuals in Study 1 on which we run our random forest models to estimate the predicted outcomes with and without treatment. As we are interested in estimating the causal effect of a cost-shock owing to switch to HDHPs in Study 2, we use the Study 2 setup to define these groups. From Study 2, we learn that the average change in cost-exposure that results from a switch to a HDHP is ~\$45 (see Figure \ref{costshock}. Based on that metric and on the nature of the data-set (distribution of costs, sample size), we define ``control" individuals in Study 1 with plan-level weighted out-of-pocket cost exposure between \$0-15 ($N_c = 55,146$), and ``treatment" individuals with plan-level weighted out-of-pocket cost exposure between \$40-70 ($N_t = 55,542$).

\subsection{Study 2: Testing for causality}
For Study 2, we focus our analysis on enrollees in our sample from the full-replacement firm and from one `comparison' firm in our data.  These `comparison' firm is one which has the same (similarly-priced) plans on offer in both the years. Including enrollees from control firm in the Study 2 analysis helps us to estimate changes in use of care attributed to the switch, as opposed to those over time.  \\ \\
We classify the enrollees in these firms based on the partitioning rules learned from Study 1 using their 2011 claims data, and then run the following difference-in-differences outcome model in each of the sub-groups on their 2011 and 2012 use of care. In Study 2, we adopt the same approach for the groups learnt from the parametric and the non-parametric approaches. We follow a negative binomial regression model.
\begin{equation}
    y_{it} = \beta_0 + \beta_1 \text{Post}_{t} \times \text{Treatment}_{i} + \beta_2 \text{Post}_{t} + \beta_3 \text{Treatment}_i + \epsilon_{it}\footnote{We cluster standard errors at enrollee level}
    \label{diffindiff}
\end{equation}
In Equation \ref{diffindiff}, Post $= 1$ in the second year (2012) and $=0$ in 2011; Treatment$=1$ for enrollees in the full-replacement firm, and $=0$ for enrollees in comparison firms. The parameter $\beta_1$ captures the difference-in-differences estimate of the effect of the switch to the high-deductible plan in each of the sub-groups.

\section{Comparison with other methods of estimating treatment effect heterogeneity}
\label{experiments}
We also evaluate how our two-study approach performs in comparison with the "causal tree" based method described in \cite{athey2016recursive}. We note that this method uses post-experiment data alone to estimate heterogeneity, while we use observational data along with experimental data. We train the ``causal" tree on a split `training' set from the experimental dataset, using the cross-validation approach as described in the study. We estimate the effects on the `validation' set from the experimental dataset, and compare the group estimates from the `training' and 'estimation' sets to determine which of the sub-groups hold up across them.

\section{Results}
\label{results}
\subsection{Findings from Study 1 Parametric approach to 2011 claims data}
\begin{figure}[H]
\centering
   \includegraphics[scale = 0.7]{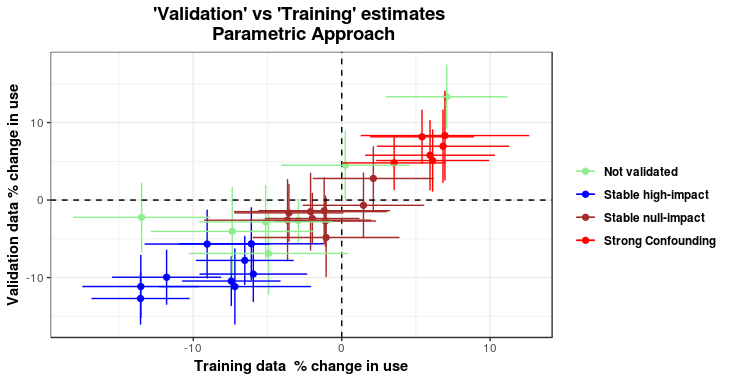}
\caption{Comparison of training and validation estimates in Study 1}	
\label{training_validation}
\end{figure}

We find 9 proposed high-impact sub-groups with persistent effects using the cross-validated model-based recursive partitioning approach. Figure \ref{training_validation} depicts the training and validation estimates (with 95 \% CI) of percent change in use of care for a \$45 increase in cost exposure for the sub-groups. 
\subsubsection{Cluster Example}
Below is an example of a proposed ``high-impact" sub-group identified in Study 1. \\ \newline \fbox{\begin{minipage}{36em}
\centering  \textcolor{blue}{{Potential 'High-impact' group}}\\
\small{Enrollees $>$ 49 yo, with $\le 6$ unique drug classe prescriptions, `medium-strata' of enrollees based on overall costs, \& claim \textbf{not} for ``Musculoskeletal System \& Connective Tissue" diseases, \\ \textbf{a \$45 $\uparrow$ in cost exposure associated with  13.5\% $\downarrow$ in OP care use}}\vspace{2ex} \end{minipage}}

\subsection{Findings from Study 1 Non-Parametric approach to 2011 claims data}
\begin{figure}[tbhp]
\centering
   \includegraphics[scale = 0.8]{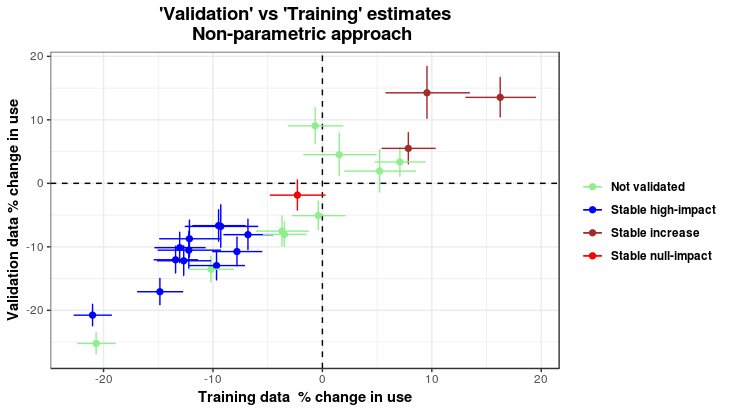}
\caption{Comparison of Study 1 non-parametric training \& validation estimates}	
\label{training_validation_np}
\end{figure}

We find 12 `high-impact' patient sub-groups with persistent effects using the cross-validated random forest approach described in Section \ref{studyonea}. Figure \ref{training_validation_np} depicts the training and validation estimates (with 95\% CI) of the proposed treatment-control difference of the cost shock. 

\subsubsection{Cluster Example from Non-parametric implementation}
Below is an  of the proposed ``stable" sub-group identified using the non-parametric approach in Study 1. \\ \newline 

\fbox{\begin{minipage}{35em}
\centering
\vspace{2ex} \textcolor{blue}{\textbf{Potential 'High-impact' sub-group}}\\
\small{Enrollees $>$ 47 yo, $\ge$ 3 unique drug class prescriptions, `medium-strata' wrt overall-cost, claim for ``Endocrine, Nutritional \& Metabolic" diseases \& \textbf{not} for `Ear, Nose, Mouth \& Throat" diseases, \\ \textbf{a $\sim$\$45 $\uparrow$ in cost exposure associated with 15\% $\downarrow$ in OP care use}}\vspace{2ex}
\end{minipage}}
\vspace{2ex} 

\subsection{Findings from Study 2 approach to experimental setting}
\label{study2findings}
Figure \ref{costshock} depicts the exogenous cost-exposure shock associated with the full-replacement firm's switch from its suite of low-deductible health offerings to high-deductible health plans. The histogram reflects the distribution of the cost exposure variable in the observational data, and we note that the cost exposure shock is of a significant magnitude (~100\%) of the pre-switch year cost exposure.
\\ \\
Table \ref{tab:causal} shows how our hypothesized `high-impact' sub-groups from the two approaches `respond' to the cost-shock due to the switch to the high-deductible plan in Study 2. We note that 4 of the 9 sub-groups we proposed to be 'high-impact' in the parametric setup were demonstrably `high-impact' in the experimental setting. Of the 12 sub-groups that we proposed to be 'high-impact' in the non-parametric setup, 7 were demonstrably `high-impact' in the experimental setting. \\\vspace{-2ex}

\begin{figure}[tbhp]
\centering
   \includegraphics[scale = 0.7]{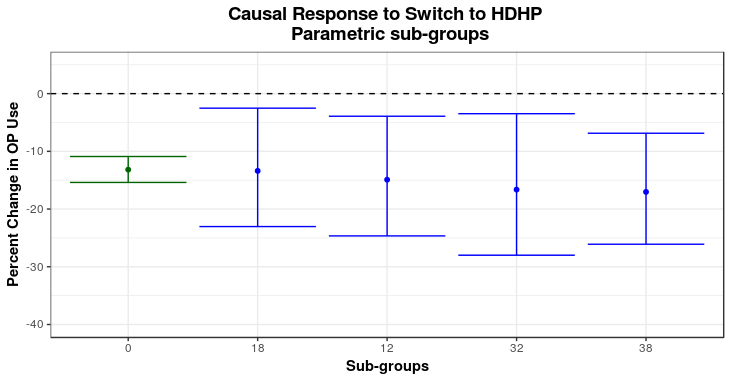} \\
   \includegraphics[scale = 0.7]{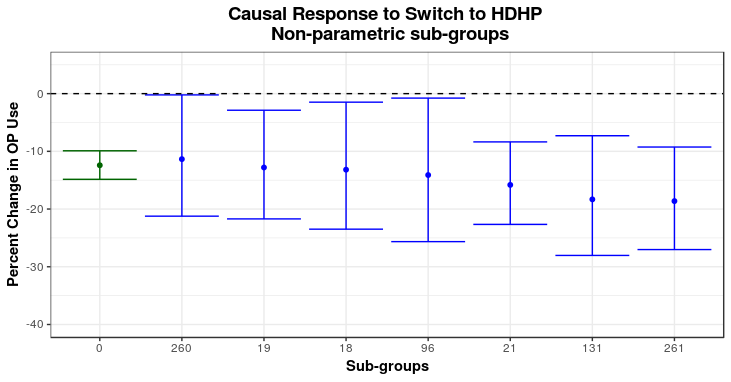}
\caption{Causal Estimates from Study 2 - Parametric and non-parametric approaches}	
\label{causal_estimates}
\end{figure}

In Figure \ref{causal_estimates}, we plot the $\beta_1$ estimates from Equation \ref{diffindiff} with the 95\% confidence interval. Cluster 0 (leftmost on the panels) is the estimate (and 95\%CI) of $\beta_1$ for the pooled data, while the remaining estimates are for clusters we hypothesized to be `high-impact' from Study 1. We note that the point estimates for our hypothesized sub-groups were on average `more negative' compared to the pooled point estimate (despite wider confidence intervals). The confidence intervals are wider for the sub-group estimates due to their relative smaller sample sizes compared to the pooled sample. 
\begin{table}[tbhp]
\centering
\begin{tabular}{l l c}
\hline
 & \textbf{Classification} &  \textbf{Clusters} \\
\hline
Parametric Method&  Confirmed in Study 2 &   $\textbf{4/9}$\\
& Fail to Reject Null &  $\textbf{5/9}$ \\
\hline 
Non-parametric Method &  Confirmed in Study 2 &  $\textbf{7/12}$\\
&  Fail to Reject Null& $\textbf{5/12}$ \\
\hline
\end{tabular}
\vspace{1em}\caption{Testing hypotheses in experimental setting (Study 2)}
\label{tab:causal}
\end{table}

\subsection{Findings from comparison with causal trees}
\begin{figure}[tbhp]
\centering
   \includegraphics[scale = 0.8]{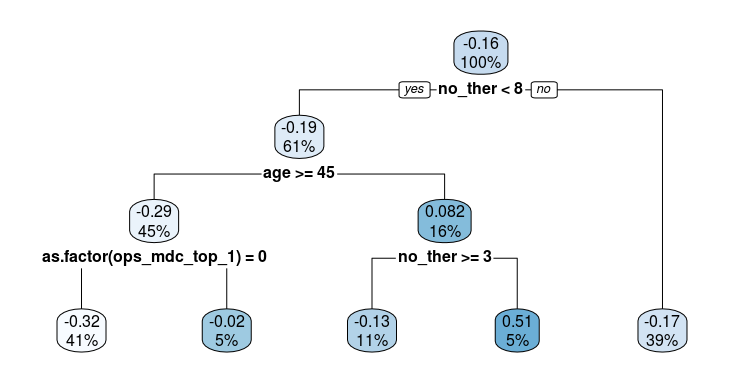}
\caption{Causal tree approach from \textcolor{blue}{\cite{athey2016recursive}} on Study 2 data}
\label{causaltree}
\end{figure}

We implemented the causal tree approach in \textcolor{blue}{\cite{athey2016recursive}}. We used only Study 2 data for this method. It identified five sub-groups on `training' set (see Figure \ref{causaltree}). \textbf{None} of the groups hold on the validation set, however. 

\section{Discussion}
\label{discussion}
Our results demonstrate that when going into an experimental study for estimating heterogeneity in treatment effects, a Study 1 approach as a precursor could be an informative step to address power limitations associated with small sample sizes. Observational data alone is limited by concerns of endogeneity which make causal inference difficult in the absence of assumptions of unobserved confounding. At the same time, the `largeness' that observational data usually affords researchers can be exploited to generate ``noise-free" or ``stable" hypotheses regarding heterogeneous treatment effects going into a subsequent experiment.
\\ \vspace{-2ex}

In our results, we note that 4 of the 9 proposed `high impact' sub-groups in the parametric approach (and 7/12 in non-parametric) were tested to be causally high-impact. These results both are informative about heterogeneity and about the nature and direction of confounding across the identified sub-groups. At the same time, the wide confidence intervals indicate that we might still be handicapped by sample size issues, and we are working on `tuning' the hyper-parameters in Study 1 (higher minimum cluster size for instance) to evaluate how that would affect our Study 2 estimates. When we use the non-parametric approach, we are able to lean 5 `high-impact' sub-groups, of which 4 hold with marginal significance. 
\subsection{Explore nature of confounding}
The results where some groups stood the causality test while some did not, indicate the need to assess the role that the direction of confounding plays in driving this. It serves to explicitly note under what conditions of confounding (same/different across groups) would hypotheses be expected to hold. This has implications based on the context in which this method can be applied. For example, in settings where we are just concerned with the ranking of effects across sub-groups and it can be assumed that all sub-groups are affected similarly by confounding, this method would find more salience than settings with differential confounding.

\subsection{Application to other datasets}
At present, we have applied this two-study approach to a setup with one full-replacement employer providing us the exogenous variation for our Study 2 setting. We are actively searching for data-sets which would fit this two-study mould. As examples, we are searching for datasets which could serve as Study 1 datasets for the Oregon Health Insurance Experiment or RAND health insurance experiment dataset.

\subsection{Continuous Treatment Regimes}
In the application to claims data, our treatment variable is the cost exposure, which is a continuous variable (notably in Study 1, with the distribution of costs across plans), while it is a discrete cost shock in Study 2 (owing to the cost shock $\delta$ happening at the plan level). To account for this, we modify  our $\gamma$ parameter to the following:
 $$\hat{\gamma}(X)= E(Y|X,A=a+\delta) - E(Y|X,A=a)$$  
We are interested in  estimating the counterfactual distribution for each of our units where $\delta$ is the cost shock (fixed as defined by Study 2) and $a$ is the level of observed cost-exposure in Study 1.

\section{Conclusion}
\label{conclusion}
When approaching an exogenous treatment intervention with an interest in heterogeneous effects, we demonstrate it is helpful to consider carrying out a Study 1 to inform it. Study 1 can use statistical machine learning methods to identify groups that may have heterogeneous treatment effects. These sub-group specific treatment effects may be biased due to confounding, but are not due to noise. Study 2 can then be used to determine which of the hypothesized effects are causal. To our knowledge, this is the first study that proposed using observational data in conjunction with experimental data in a two-study approach to estimate heterogeneity.
\\ \vspace{-2ex}

As the results demonstrate, further exploration of the direction of confounding in both the `low impact' and `high impact' sub-groups is called for. As future work, we also plan to incorporate medical and clinical review for interpretability of these clusters and to better understand mechanisms for heterogeneous causal effects based on these partitioning variables. Methodologically, future work will involve applying the proposed non-parametric approach to sub-group estimation and explorations of applying this framework to dynamic treatment regimes. 

\bibliography{references}
\end{document}